\documentclass{optica-article}

\journal{opticajournal} % for journals or Optica Open

\articletype{Research Article}

\usepackage{lineno}
\usepackage{comment}
\usepackage{orcidlink}
\usepackage{subfig}
\usepackage{svg}
\usepackage[utf8]{inputenc}

\DeclareMathOperator{\sinc}{sinc}
\DeclareMathOperator{\tr}{tr}

\begin{document}

\title{Impact of polarization mode dispersion on entangled photon distribution}

\author{Vadim Rodimin\authormark{1, *}\orcidlink{0000-0003-2982-7768}, Konstantin Kravtsov\authormark{1}\orcidlink{0000-0003-4499-4089}, Rui Ming Chua\authormark{1,2}\orcidlink{0000-0002-9814-9190}, Gianluca De Santis \authormark{1}\orcidlink{0009-0005-2393-6076}, Aleksei Ponasenko \authormark{1}\orcidlink{0009-0007-7529-6415}, Yury Kurochkin\authormark{1}, Alexander Ling\authormark{2,3}\orcidlink{0000-0001-5866-1141}  and James A. Grieve\authormark{1}\orcidlink{0000-0002-2800-8317}}

\address{\authormark{1}Quantum Research Center, Technology Innovation Institute, Masdar City, Abu Dhabi, UAE\\
\authormark{2}Centre for Quantum Technologies, 3 Science Drive 2, National University of Singapore, 117543 Singapore\\
\authormark{3}Department of Physics, National University of Singapore, Blk S12, 2 Science Drive 3, 117551 Singapore}

\email{\authormark{*}vadim.rodimin@tii.ae} %% email address is required; see note below about the corresponding author designationYury Kurochkin

% use {asbstract*} to suppress the copyright line. Copyright information will be added in production

\begin{abstract*}
%%%VR%%%
The polarization mode dispersion (PMD) in optical fibers poses a major challenge for maintaining the fidelity of quantum states for quantum communications. In this work, a comprehensive model linking the probability of quantum measurement errors (infidelity) to PMD is developed and validated by experimental measurements of differential group delay and quantum bit error rate (QBER). Our research proposes effective methods to mitigate PMD effects for broadband entangled photons and evaluates the impact of higher-order PMD effects. The model provides an experimentally verified framework for the optimization of commercial quantum key distribution systems in deployed fiber optic lines.
\end{abstract*}

%%%%%%%%%%%%%%%%%%%%%%%%%%  body  %%%%%%%%%%%%%%%%%%%%%%%%%%
\section{Introduction}

%%% VR %%%
Quantum communication is a rapidly developing field that focuses on facilitating communication between quantum computers~\cite{Knaut2024-bw, Togan2010-ob, Nemoto2016-vh}, or applying the principles of quantum mechanics to cryptography in the technology of  Quantum Key Distribution (QKD)~\cite{ekert1991quantum, BBM92Original, PhysQuantInfo}. Like its classical counterpart, the communication process is often facilitated by fiber optics, particularly within Metropolitan Area Networks~\cite{Chen2021}. However, fiber optics has its unique brand of challenge to communications, where phenomena such as chromatic and polarization mode dispersion diminish the efficacy of the communications process and thus have to be managed~\cite{Gallager_2008}.

Polarization Mode Dispersion (PMD) occurs in optical fibers, causing wavelength-dependent polarization transformations of light as it travels through the fiber. This phenomenon results from the residual birefringence in the fiber induced by the manufacturing imperfections and stress. In an ideal single-mode fiber, light propagates without any depolarization, while in the real-world situation we observe varying propagation speeds for different polarization modes at different wavelengths \cite{Galtarossa2005}.

\begin{figure}[ht!]
\centering\includegraphics[width = 0.45\textwidth]{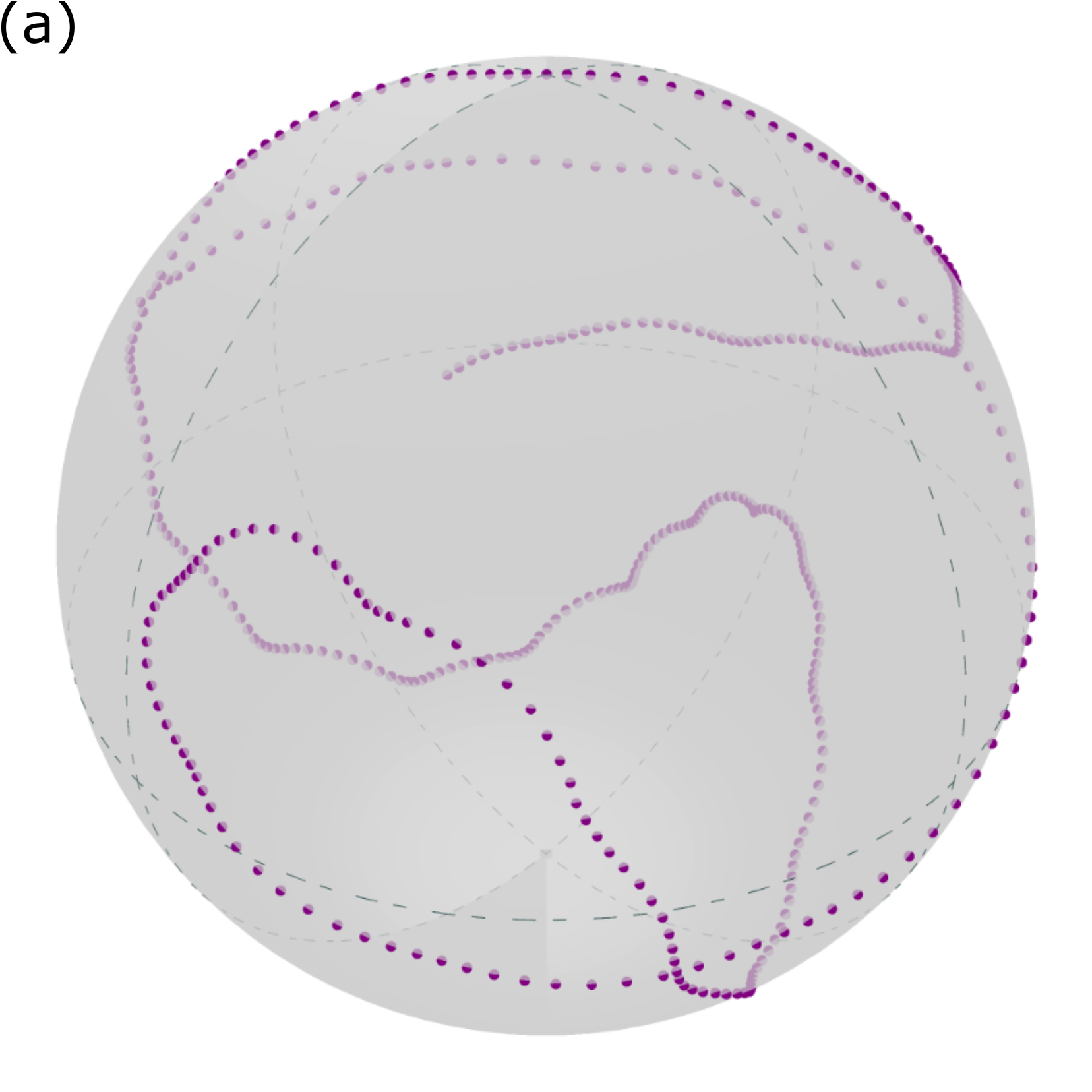}
\centering\includegraphics[width = 0.48\textwidth]{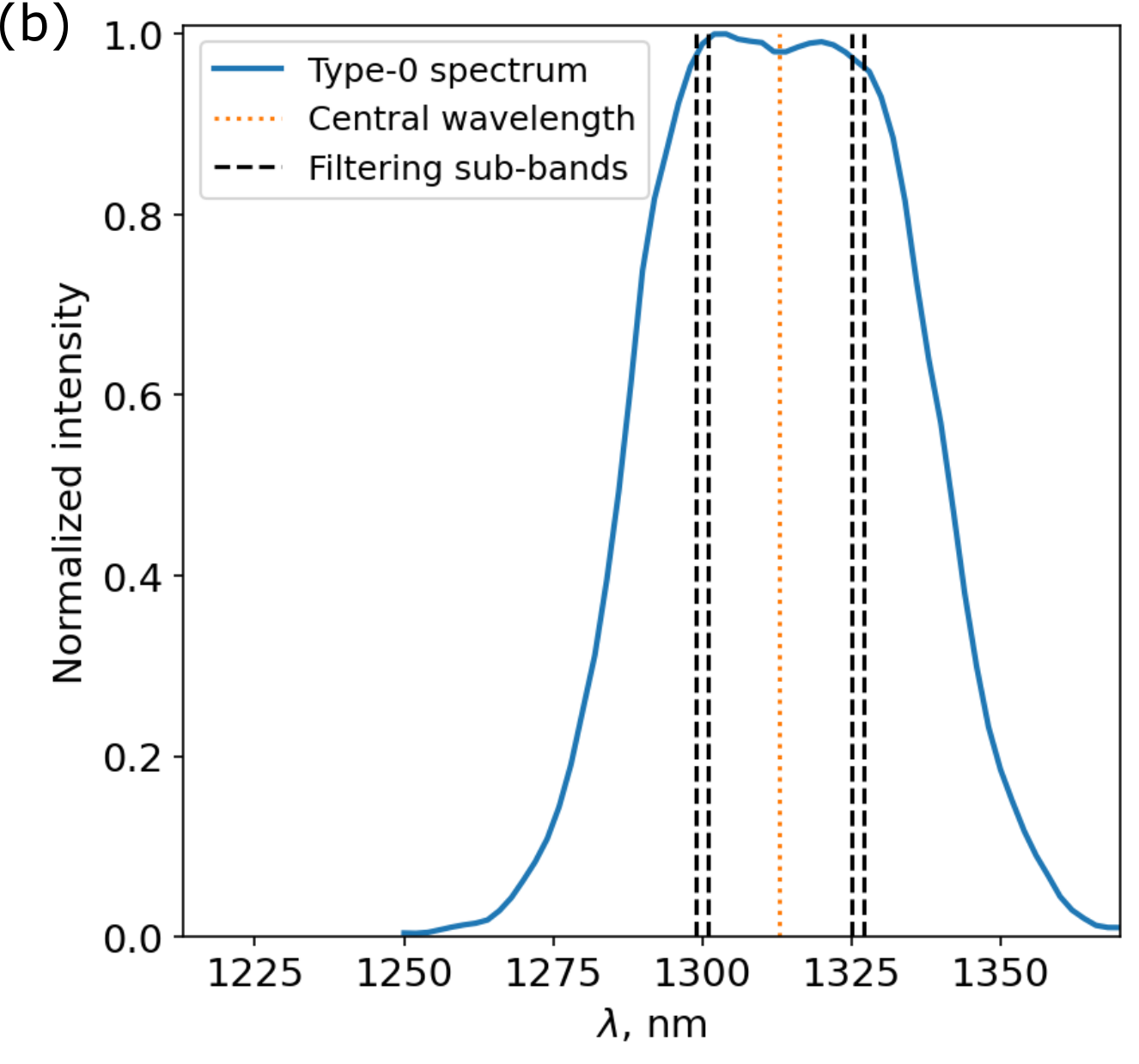}
\caption{Necessity of narrow sub-bands filtering due to PMD. (a) Observed polarization trajectory upon propagation through 30 km of typical telecom fiber, measured over the range of 1260 - 1360~nm in steps of 0.25~nm; (b) Example of a broadband type-0 SPDC spectrum, produced by 25~mm of Periodically Poled Lithium Niobate (PPLN) crystal with the poling period of 12.83~$\mu m$. The pumping wavelength is 656.5~nm.}
\label{fig:spectrum}
\end{figure}

In this work, we use a graphical approach, visualizing the state of light on the Poincaré sphere. PMD in single-mode fibers manifests itself as a change of the output polarization state with the change of the wavelength while the input polarization remains constant.  Figure~\ref{fig:spectrum}a. shows an exemplary output polarization trajectory on the surface of the Poincaré sphere measured for an installed 30~km long fiber over a 100~nm bandwidth. 

If the polarization degree of freedom is used for classical communications it is sufficient to use two orthogonally polarized states. They may be chosen as the Principal States of Polarization (PSPs), which have minimal depolarization.
In the context of quantum communications the use of just two orthogonal polarization states is insufficient. Thus, the challenges brought upon by the PMD phenomena are more pronounced. First, one must use superpositions of orthogonally polarized states \cite{BENNETT20147} in order to achieve an advantage over classical communications.

Second,
practical generation of non-classical light, such as Spontaneous Parametric Down Conversion (SPDC), often results in a much more broadband signals than generated by conventional laser sources. It is especially the case for the most efficient type-0 process known for its relatively broad spectrum --- see Fig.~\ref{fig:spectrum}b. This also exacerbates the detrimental effects of the PMD [Review paper on SPDC]. Thus, maintaining the fidelity of quantum states becomes a significant challenge that directly affect the quality of QKD and other applications of quantum communications.

%the detrimental effects of dispersion go beyond Inter Symbol Interference, and a reduction in rate -- decoherence can disrupt the success of the entire process [Robert G. Gallager, Cite relevant papers on decoherence]. 

In our study, we distribute bi-partite entangled states 
\begin{equation} \label{eq:state}
|\psi\rangle =
 \frac{|HH\rangle + |VV\rangle}{\sqrt{2}},
\end{equation}
defined in terms of the H (horizontal) and V (vertical) basis states. Each of the two photons is essentially in the fully mixed state, while their measurements in the same bases yield strong correlations.  We assume that the distributed entanglement is later used in BBM92 QKD protocol \cite{ekert1991quantum, BBM92Original}, however, the main results are directly applicable to other possible entanglement utilization.

%One of the most effective entanglement sources in the optical domain is type-0 Spontaneous Parametric Down-Conversion (SPDC), known for its relatively broad spectrum Fig.~\ref{fig:spectrum}b. Factors such as crystal length, pump bandwidth, focusing conditions, and temperature influence this spectrum. 

 The most direct way of decreasing PMD-induced distortions and improving correlations is the reduction of the spectral width of the source using narrowband filtering before transmitting the light into the fiber. Fig.~\ref{fig:spectrum}b shows an example of filtering distant wavelengths, which is convenient for separating entangled photons by channels. However, such filtering directly affects the brightness of the source, creating a trade-off between the brightness of the transmitted light and the observed quantum bit error rate (QBER). Optimization of the system, thus, requires a model linking the channel parameters and the bandwidth of the signal with the level of distortion experienced in BBM92-type measurements. The most common parameter describing the PMD in a fiber channel is the Differential Group Delay (DGD) \cite{Galtarossa2005} -- the difference in transit time for light launched into two PSP modes. Our main goal is finding a simple relation, connecting the measurement error probability with the filtering bandwidth and the DGD of the channel.

Effects of various types of optical dispersion in fibers on QKD applications present several challenges and have been studied previously. Nonlocal dispersion compensation has been proposed \cite{franson1992nonlocal} and experimentally validated \cite{neumann2021experimentally,chua2022fine} to address the chromatic dispersion.
Polarization mode dispersion was also extensively studied in \cite{Antonelli2011, Brodsky2011, Lim2016, Riccardi2021} providing significant insights. However, we didn't find a concise way to apply them to derive a measurement errors vs. DGD expression. Some of the previous studies, such as \cite{Xavier2008}, directly focus on prepare-and-measure protocols but are hardly applicable to broadband entangled QKD. The model proposed in \cite{Muga2011} provides a practical framework that aligns well with our objectives; however, its final results lack experimental validation and do not satisfactorily correspond to our findings.

The paper is organized as follows. In section 2 we present a graphical approach linking the measurement error probability with the PMD. Then Section 3 provides experimental validation through DGD measurements and QBER analysis in a real QKD system. Finally, in the Discussion section, we propose methods to mitigate PMD effects for broad-band entangled-based QKD and assess the impact of higher-order PMD effects.

\section{Model description}
%The next 2 commented paragraphs came from the Introduction.

%The photon source creates an entangled state; each components are uniformly distributed across the Poincaré sphere. At the receivers' locations, the states are measured in two mutually unbiased measurement bases, for example horizontal-vertical (H-V) and diagonal-antidiagonal (D-A). These states are aligned on a great circle of the Poincaré sphere, so we define a 4-state circle as a great circle on the Poincare sphere comprising of the measurement basis vectors. As these states propagate through a fiber, they undergo wavelength-dependent PMD, which fluctuates over time due to environmental changes.

%Studies on PMD classify the effect of PMD into first-order PMD and higher-order PMD \cite{DGDLeNelsonIntroduction1stOrder2ndOrder}. First-order PMD considers a constant relative propagation delay between the two Principal States of Polarization (PSPs). In contrast, higher-order PMD involves variations in both the direction of the PSP and the delay value with wavelength. For two orthogonal polarization states, the effect of first-order PMD can be mitigated by aligning the input polarization with the PSPs of the fiber. However, for the previously mentioned 4-state circle, significant first-order PMD effects cannot be eliminated in the same manner.

\begin{figure}[ht!]
\centering\includegraphics[width = 0.40\textwidth]{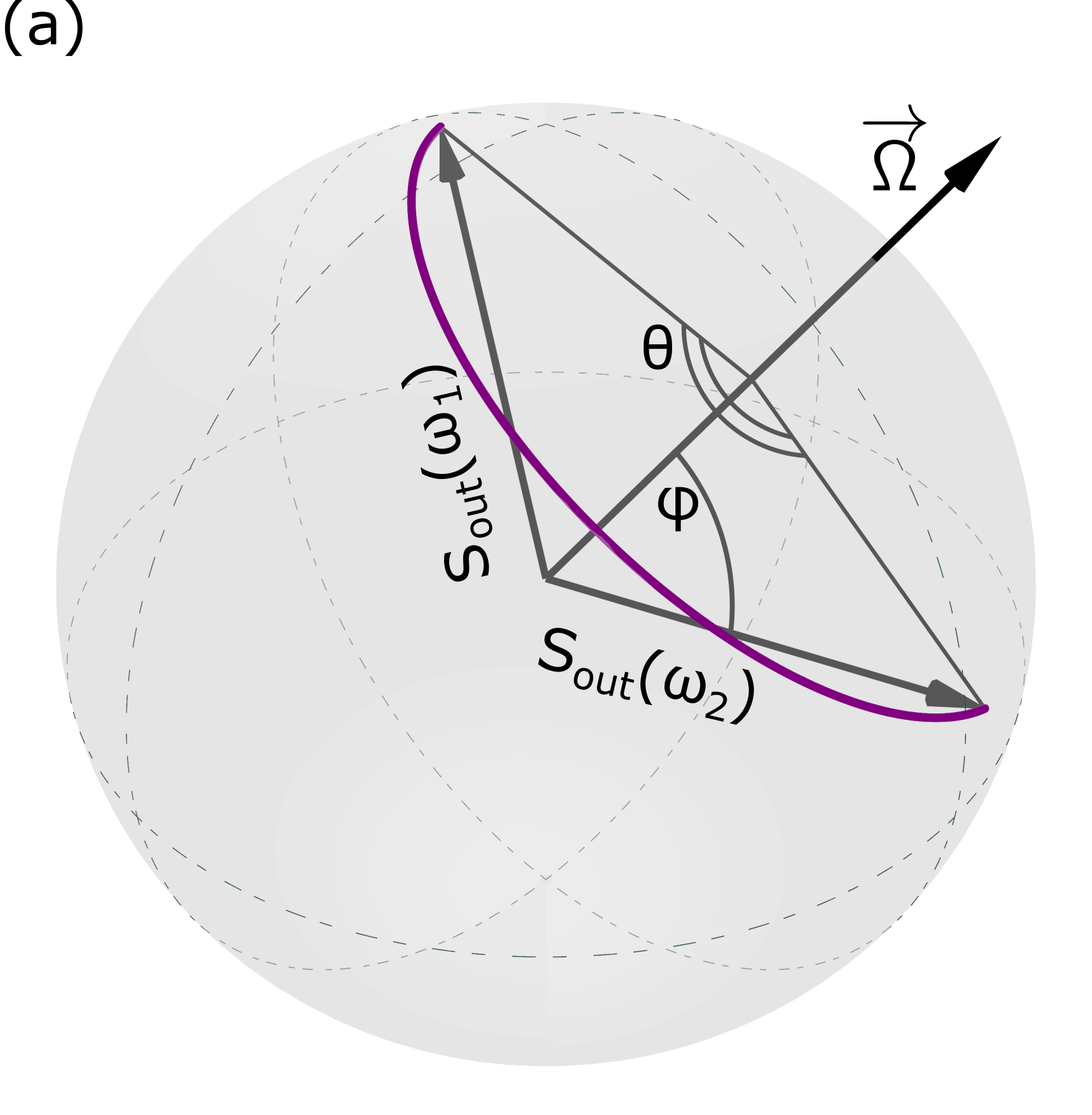}
\centering\includegraphics[width = 0.40\textwidth]{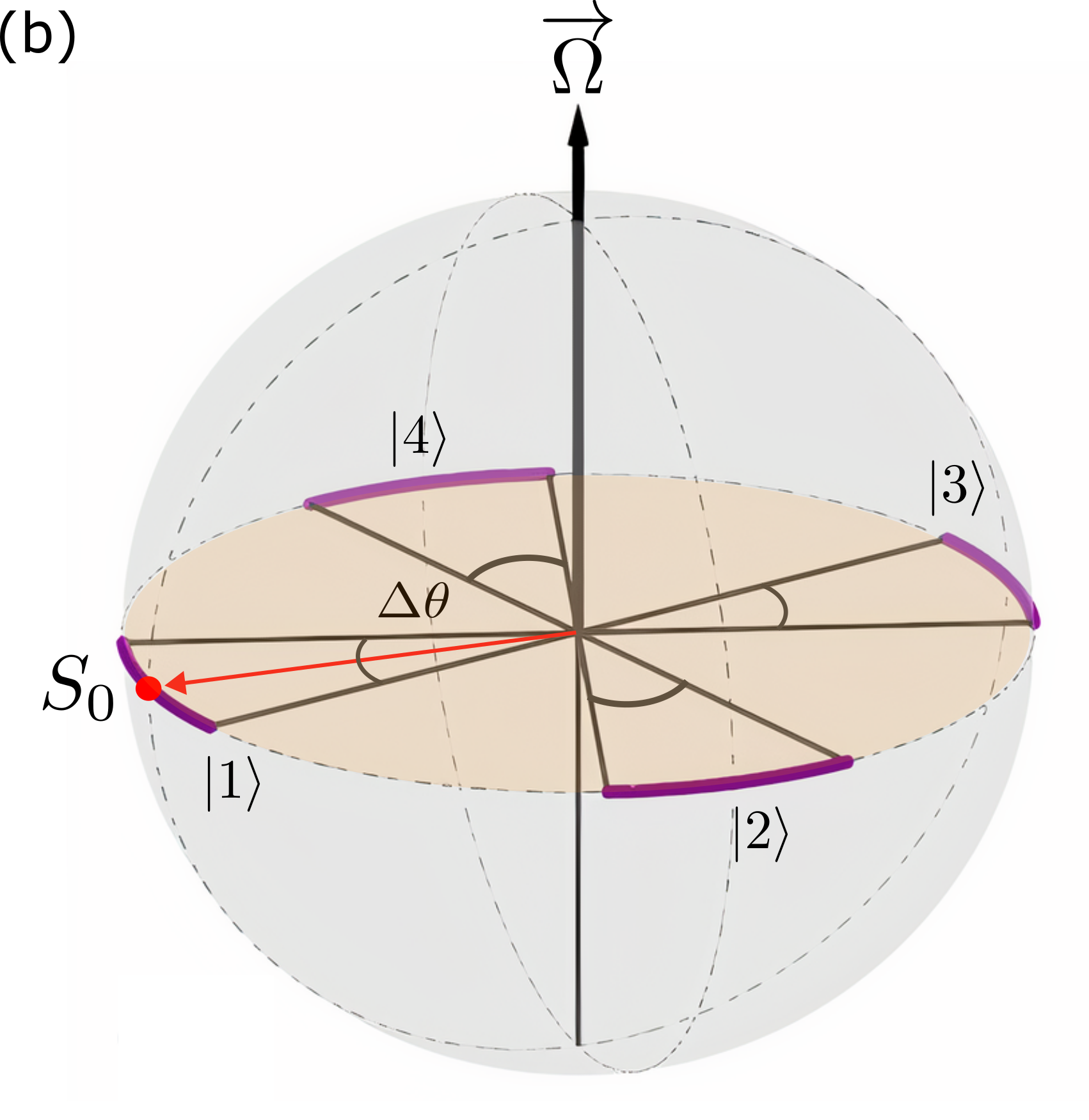}
\centering\includegraphics[width = 0.40\textwidth]{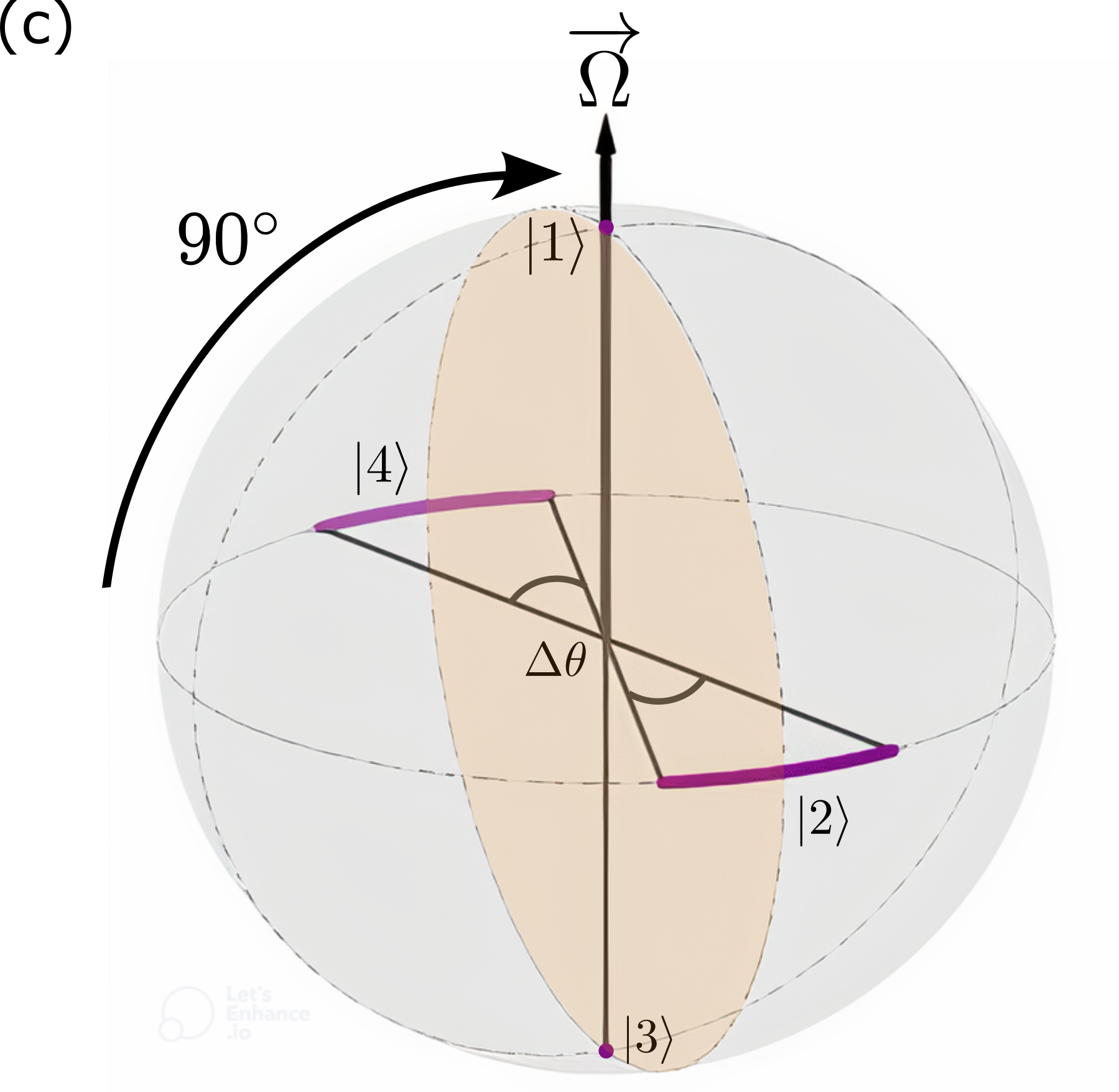}
\centering\includegraphics[width = 0.40\textwidth]{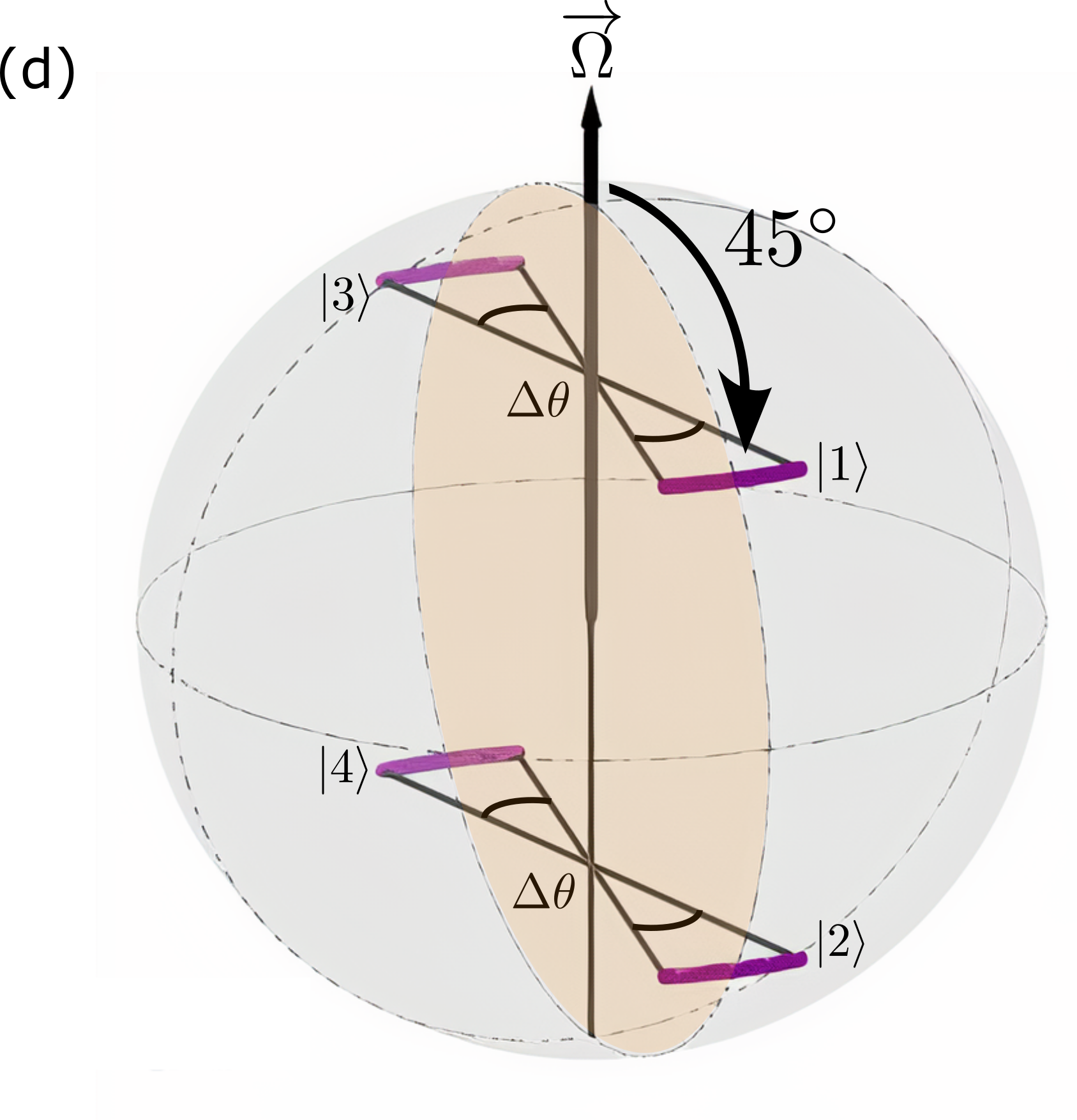}
\caption{Polarization trajectories on the Poincare sphere with wavelength variation. (a) the first-order PMD: output polarization rotates around the PMD vector; (b) the largest distortion case, where the PMD vector is orthogonal to 4-state circle plane; (c) the PMD vector lies in the 4-state circle plane coinciding with one of the basis, in which no distortions are observed; (d) a symmetrized case, where the PMD vector lies in the 4-state circle plane.}
\label{fig:spheres}
\end{figure}

The transformation of the polarization state $s_{out}$ in a fiber may be described by the PMD vector $\vec{\Omega} = \Delta\tau \Vec{p}_1$, where $\Delta\tau$ and $\Vec{p}_1$ are DGD and PSP vector, respectively~\cite{DGDLeNelsonIntroduction1stOrder2ndOrder}. When the light frequency changes from $\omega_1$ to $\omega_2$, the output polarization state rotates around the PMD vector, as shown in Fig.~\ref{fig:spheres}a. The first order PMD is defined by the linear relation 
\begin{equation} \label{eq:rotation}
\Delta\theta = \Delta\tau \cdot (\omega_2 - \omega_1)
\end{equation}
between the rotation angle $\Delta\theta$ and the DGD, where the PMD vector is assumed to remain constant across all frequencies in the given range.
On the contrary, the higher orders of PMD define variations of the PMD vector's length and orientation with changes in frequency \cite{DGDLeNelsonIntroduction1stOrder2ndOrder}. 
Thus, the 1-st order PMD describes a regular circular rotation around $\vec{\Omega}$ on the Poincaré sphere. The actual trajectory of polarization transformation from Fig.~\ref{fig:spectrum}a can be matched by assuming frequency-dependent parameters $\Delta\tau(\omega)$, $\Vec{p}_1(\omega)$, and resulting $\vec{\Omega}(\omega)$~\cite{Shtaif2004}. Longer trajectories of an output polarization state on the Poincaré sphere correspond to greater distortions caused by the PMD.

In the context of the BBM92 protocol, the photon source creates an entangled state~(\ref{eq:state}), where each component is uniformly distributed across the Poincaré sphere. At the receivers' locations, the components are randomly measured in one of the two mutually unbiased bases, for example, in horizontal-vertical (H-V) and diagonal-antidiagonal (D-A). These four states define a great circle on the Poincaré sphere, which we will call below a 4-state circle. Propagation over the fiber effectively modifies these basis states including the effects of the wavelength-dependent PMD. All polarization changes in the fiber are subject to fluctuations over time due to environmental changes.

 For simplicity, we are starting with the 1-st order PMD, which is an accurate approximation if we consider small enough trajectory patches.  An important consideration is that the 4-state circle, formed by the measurement basis vectors, can have different orientations with respect to the PMD vector. The longest total trajectory corresponds to the PMD vector oriented perpendicularly to the 4-state circle (Fig.~\ref{fig:spheres}b). However, it is also possible to reduce the PMD-induced distortion by choosing the measurement bases such that the 4-state circle plane comprises the PMD vector. If a basis coincides with the PSP vectors, the corresponding trajectories reduce to single points, so only the opposite basis contributes to PMD-caused measurement errors as shown in Fig.~\ref{fig:spheres}c. Another notable configuration is a symmetrical one, depicted in Fig.~\ref{fig:spheres}d, where the PMD vector lies in the 4-state circle plane between two adjacent basis states.

Consider a trajectory $L$ on the Poincaré sphere created by broadband light with a spectrum evenly distributed within a bandwidth $\Delta\lambda$. When measuring the output polarization projecting it on a chosen state $|s_0 \rangle$, the intensity is directly proportional to the integral of %the squared modulus of the projections 
the corresponding amplitude modulus squared over all polarization states within the trajectory $L$:

\begin{equation} \label{eq:intensity}
%I \propto \int_L |\langle s(\lambda)s_0 \rangle|^2 d\lambda
I \propto \int_L \bigl|\langle s(\lambda) | s_0 \rangle\bigr|^2\, d\lambda %%% KK %%%
\end{equation}

In the case of single-photon measurements in the basis $\{s_0, s_{0 \mathrm{\_orthogonal}}\}$, the measurement error probability $p_e$ can be determined using

\begin{equation} \label{eq:arc_int}
p_e = 1 - \frac{1}{\Delta\lambda_L} \int_L \bigl|\langle s(\lambda)|s_0 \rangle\bigr|^2\, d\lambda, %%% KK %%%
\end{equation}
which we refer below as a "trajectory integration". It can be understood as the PMD contribution to QBER for a particular state measurement.
%%% KK %%%%
It is easy to show that $p_e$ is in fact the {\em infidelity} \cite{Jiang2020} between the median pure state $\rho_0 = |s_0\rangle\langle s_0| $ and the resulting density matrix
\begin{equation}
\rho(\Delta\lambda) = \frac{1}{\Delta\lambda_L} \int_L |s(\lambda)\rangle\langle s(\lambda)| \, d\lambda,
\end{equation}
where infidelity is defined as $1-F\bigl(\rho_0, \rho(\Delta\lambda)\bigr) = 1 - \Bigl(\tr \sqrt{\rho_0 \rho(\Delta\lambda) }\Bigr)^2$ \cite{baldwin2023fidelity}.

Now we consider a relatively small arc on the circle formed by the angle $\Delta\theta$ corresponding to the filtering bandwidth $\Delta\lambda$.
%We can tune the actual measurement basis with a polarization controller, so any $s_0$ can be chosen for the projective measurement. 
Given a uniform spectral distribution, we choose $s_0$ be the central point of the arc in Fig. \ref{fig:spheres}b, so the overall error probability is minimized. 
The resulting infidelity is obtained through integration over all wavelengths

\begin{equation} \label{eq:model}
%p_e = 1 - \frac{2}{\Delta\theta} \int_0^{\Delta\theta/2} \cos^2\frac{\theta}{2} d\theta   \approx  \frac{(\Delta\theta)^2}{48},
p_e = 1 - \frac{2}{\Delta\theta} \int_0^{\Delta\theta/2} \left(1- \sin^2\varphi \sin^2\frac{\theta}{2}\right)\,d\theta  = \frac{\sin^2 \varphi}2
\left[ 1-\sinc \frac{\Delta\theta}2\right], %%% KK %%%
\end{equation}
where $\varphi$ is an angle between the polarization state and the PMD vector $\vec{\Omega}$ (Fig.~\ref{fig:spheres}a).
It has to be mentioned here that the $\sinc$ shape of the infidelity is effectively the Fourier transform of the assumed rectangular signal spectrum. Gaussian spectral distribution would result in a Gaussian infidelity shape, and so on.

For the most practically interesting case of relatively small $p_e$ this can be further simplified using Taylor expansion in powers of $\Delta\theta$, to get a useful relation %%% KK %%%
\begin{equation} \label{eq:modelsimple}
p_e \approx \sin^2 \varphi \, \frac{(\Delta\theta)^2}{48} %%% KK %%%
\end{equation}

\begin{figure}[ht!]
\centering\includegraphics[width = 0.48\textwidth]{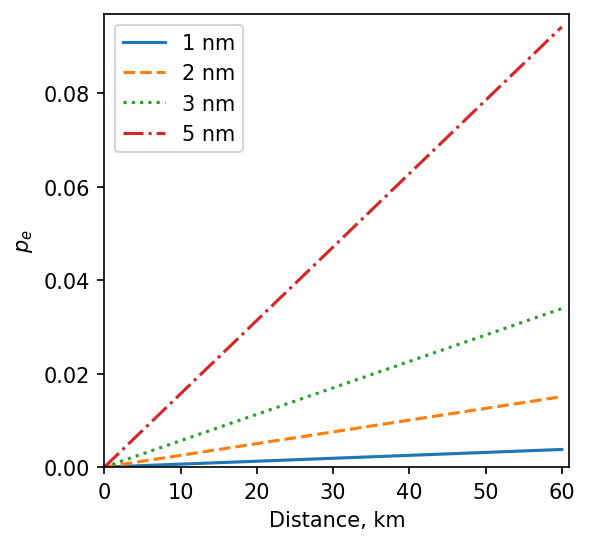}
\centering\includegraphics[width = 0.48\textwidth]{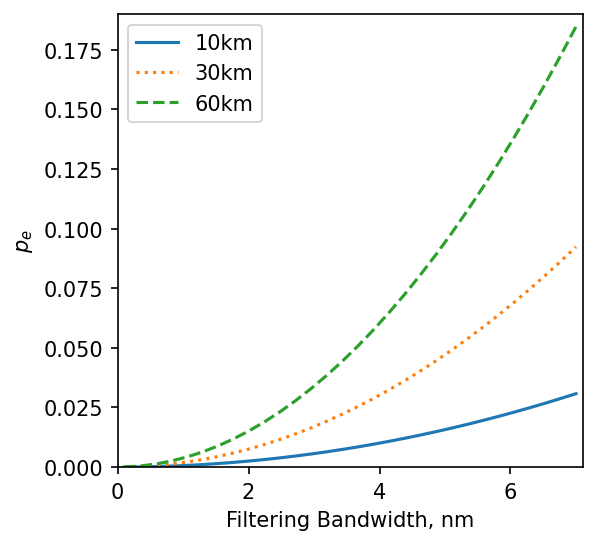}
\caption{Expected infidelity $p_e$ vs. fiber link distance (left) and the bandwidth of filtering (right). For the calculation we used a typical PMD link value of 0.05 ps$/\sqrt{\mathrm{km}}.$}
\label{fig:mean_stat}
\end{figure}

From a known DGD value we can find $\Delta\theta$ using eq.~(\ref{eq:rotation}) that is now directly related to the expected $p_e$. As a result, we connect the effective infidelity with the filtering bandwidth and the QKD distance, with some examples shown in Fig.~\ref{fig:mean_stat}. The relationship between the propagation distance and $p_e$ is linear, because the DGD value scales as a square root of the distance. The dependence on the bandwidth exhibits a quadratic behavior.

\section{Experimental validation}
To estimate the actual PMD-induces signal degradation in a real-life scenario we exploit two different methods.

\textbf{First Principles Approach}:
This estimation is based on the actually measured polarization trajectory on the Poincaré sphere under varying input wavelength. For a certain signal bandwidth $\Delta \lambda$ we find the corresponding trajectory patch on the Poincaré sphere. Assuming the optimal alignment of the measurement basis, we choose the measurement state $s_0$ as the central point of this patch. As one can see in Fig. \ref{fig:spheres}b, the measurement $S_0$ is affected by the PMD distortion of the original state $|1\rangle$. Infidelity is calculated through numerical integration (\ref{eq:arc_int}) along the patch. For a longer trajectory, as illustrated in Fig. \ref{fig:spectrum}a, the patch is rolled along the trajectory, and the rolling integration yields infidelity as a function of wavelength for a given $\Delta \lambda$.

\textbf{DGD-Based Approach}:
This method involves measuring the DGD as a function of wavelength for a given fiber channel, using standard telecom techniques. Using the derived formulas (\ref{eq:rotation}, \ref{eq:modelsimple}), we then calculate the infidelity as a function of the wavelength for the specified filtering bandwidth, $\Delta \lambda$ ($\Delta \omega$).

In this section, we verify the consistency between these two methods and compare the obtained results with those observed in a real QKD system.

\begin{figure}[ht!]
\centering\includegraphics[width = 0.7\textwidth]{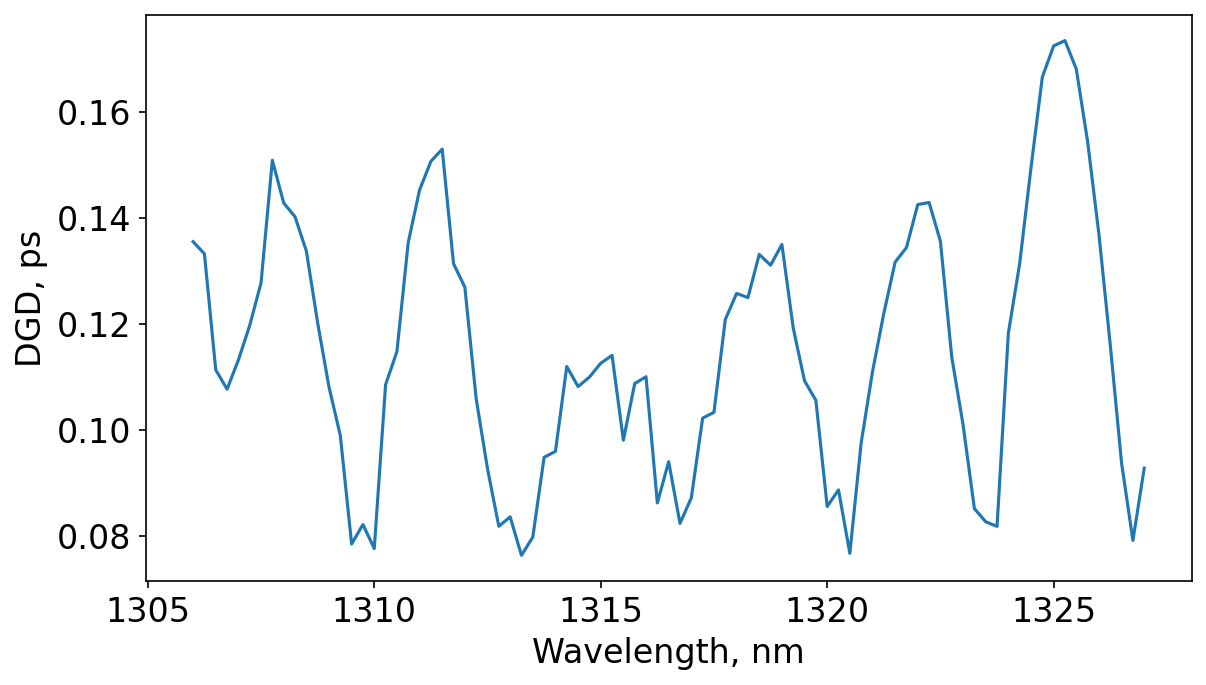}
\caption{The measured differential group delay vs. wavelength obtained using MMM~\cite{Jopson1999} method.}
\label{fig:DGD}
\end{figure}

In order to measure the DGD we rely upon the frequency-domain measurement setup \cite{PaulWilliamsPMDExperimentMullerMatrix}. It consists of connected in-series a tunable laser, a polarization controller, the fiber span itself, and a polarimeter, see Supplementary Materials (SM), Fig. S1. We carried out polarization measurements of various 10~km deployed fiber channels in Masdar City, Abu Dhabi. We used the Mueller Matrix Method (MMM) \cite{Jopson1999}, to calculate the DGD as a function of the wavelength. The results of DGD measurements are shown in Fig.~\ref{fig:DGD}, see SM, Section 2 for more details.

\begin{figure}[ht!]
\centering\includegraphics[width = 0.8\textwidth]{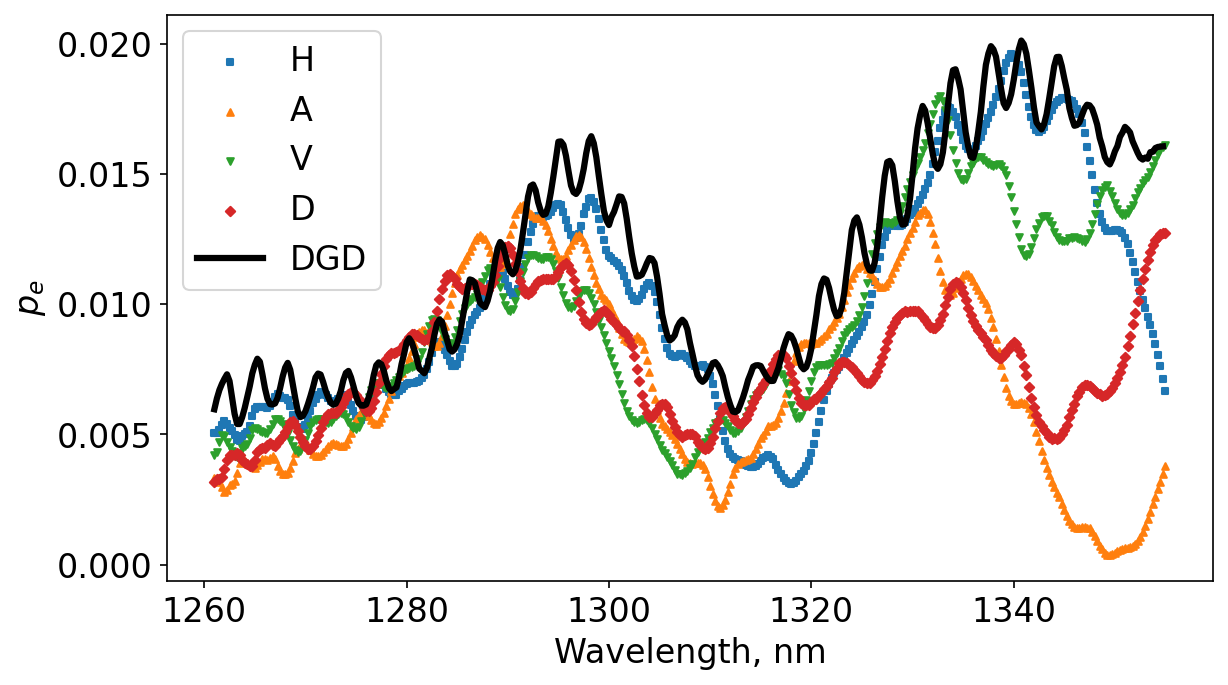}
\caption{Infidelity $p_e$ for a 5~nm filtering bandwidth in a 10~km deployed fiber loop. Solid black: DGD-based estimation using (\ref{eq:modelsimple}). Other colors: trajectory integration (\ref{eq:arc_int}) for 4 different input polarization states H, V, D, and A.}
\label{fig:dQBER_vs_lambda}
\end{figure}

In Fig.~\ref{fig:dQBER_vs_lambda} we compare the expected infidelity values obtained with these two methods for a fixed bandwidth of 5~nm. The infidelity values obtained from numerical integration are smaller than those derived using the DGD-based method. We attribute this difference to the following considerations. An arbitrary polarization generally does not rotate around the PMD vector along the great circle. A random state can rotate at any $\varphi<\pi/2$ as shown in Fig.~\ref{fig:spheres}d or even converge to a point as for states $|1\rangle$ and $|3\rangle$ in Fig.~\ref{fig:spheres}c. When we find infidelity from DGD measurement (\ref{eq:modelsimple}), we assume the worst case of $\sin\varphi = 1$, corresponding to the rotation along the great circle in Fig.~\ref{fig:spheres}b. Consequently, the DGD-based infidelity represents an upper limit for the trajectory integration method. Therefore, comparing DGD calculation with trajectory integration validates the proposed model. The slight deviation of the DGD (\ref{eq:arc_int}) curve from the maximum possible can be explained by neglecting the higher-order PMD effects in the simplified model (\ref{eq:modelsimple}).

\begin{figure}[ht!]
\centering\includegraphics[width=0.8\textwidth]{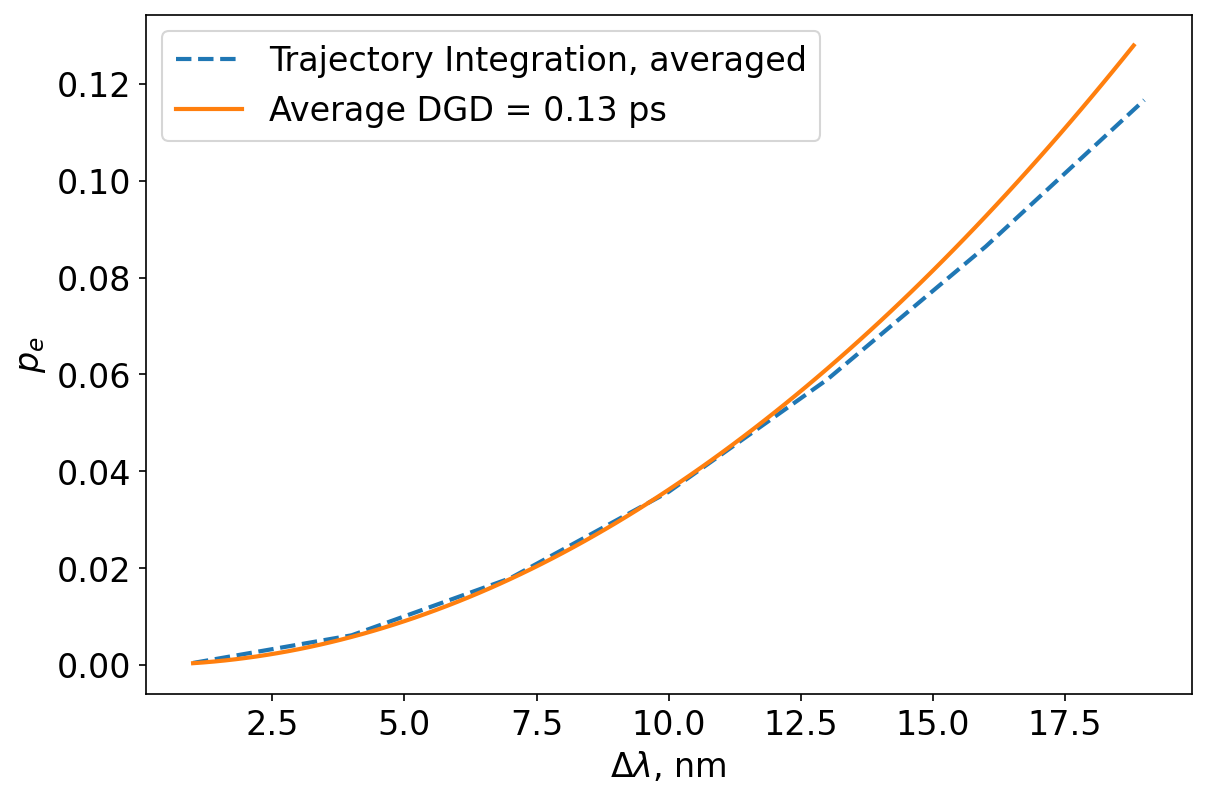}
\caption{Comparison of infidelity calculation using the DGD-based approach (\ref{eq:modelsimple}) and the numerical trajectory integration (\ref{eq:arc_int}) for a varying filtering bandwidth $\Delta\lambda$.}
\label{fig:dQBER_vs_filtering}
\end{figure}

In Fig.~\ref{fig:dQBER_vs_filtering}, we assess the correspondence between the two methods for estimating $p_e$%: numerical trajectory integration (\ref{eq:arc_int}) and proportionality to DGD squared (\ref{eq:modelsimple}) while
while altering the filtering bandwidth $\Delta\lambda$. The relationship is quadratic and aligns closely with the proposed model.

\begin{figure}[ht!]
\centering\includegraphics[width=0.8\textwidth]{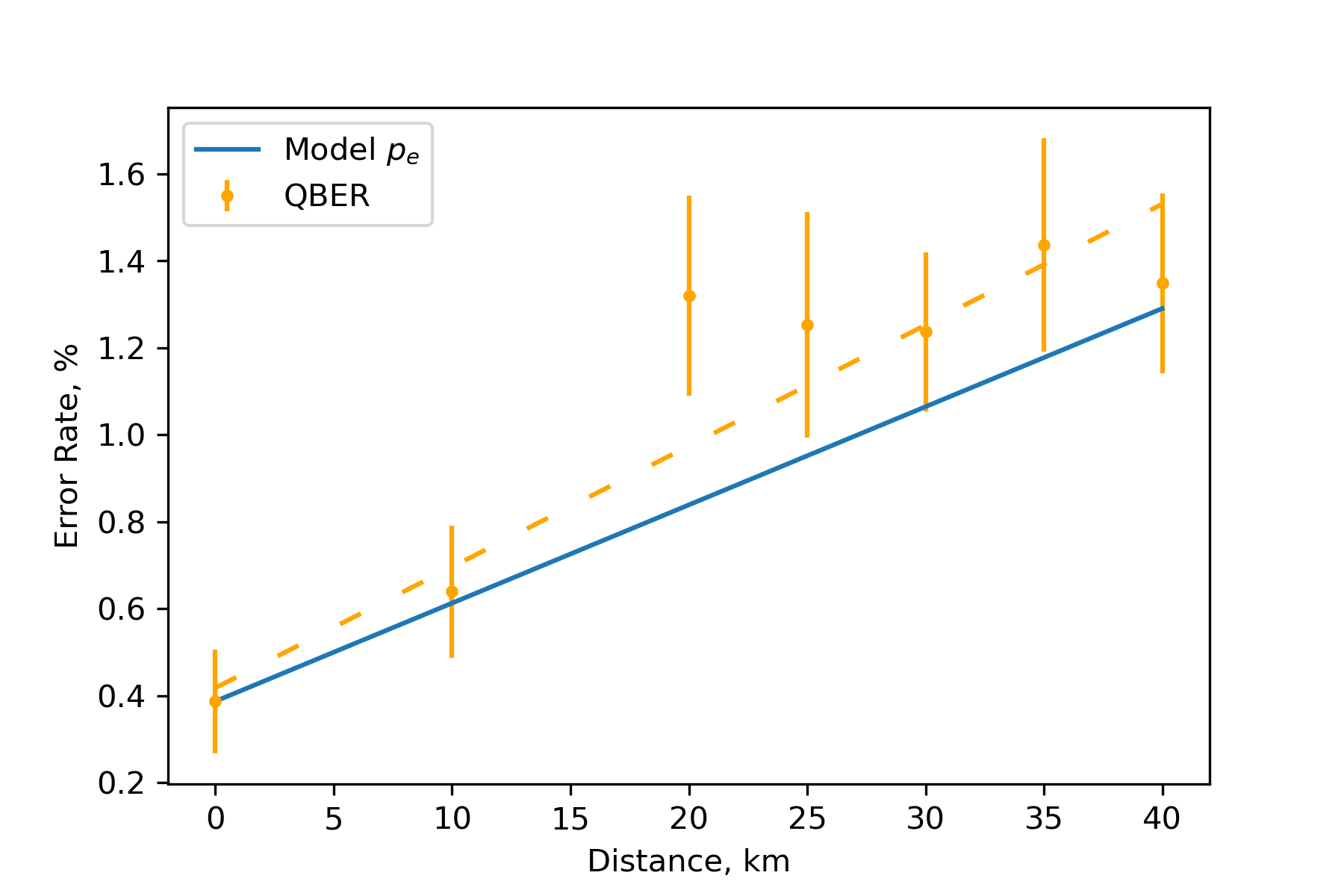}
\caption{Measured QBER and modeled infidelity $p_e$ vs. link distance, PMD link value $= 0.05 ps/\sqrt{km}$, typical for our fibers. The dashed line represents the linear regression of the experimental QBER points and is provided as a visual guide. The modeled line is adjusted upward to correspond to a zero-distance QBER.}
\label{fig:QKD_Results}
\end{figure}

Despite our focus on PMD effects, in real QKD experiments it is impossible to isolate the PMD-induced infidelity from other contributions, such as background counts and basis misalignments. However, for short distances and relatively broad filtering bandwidths, we can reasonably expect PMD-induced errors to dominate over other factors. Moreover, the total error rate should at least exceed the PMD-related predictions for the system. To verify the consistency of our model with these assumptions, we conducted a series of experiments using a self-made QKD setup prepared for real-world deployment. The setup is based on the same scheme used in~\cite{Shi2020}, while its detailed description is given in the SM, Section 3.

As a source of polarization-entangled photon pairs our setup uses a PPLN crystal within a linear displacement interferometer, applying a technique similar to that described in \cite{EntangledPhotonPairSourceLohrmann}. The PPLN crystal generates an SPDC spectrum centered at 1313 nm with an FWHM range of 30–100 nm.  Factors such as crystal length, pump bandwidth, focusing conditions, and temperature influence the resulting spectrum. The photon pairs are separated into two channels using fiber Bragg filters with 2 nm sub-bands, while fiber spools were utilized to make up the distance. 

For on-site detection, we used a passive basis choice optical scheme with IDCube single-photon detectors set to 20$\%$ detection efficiency. An automated polarization controller in one of the channels is employed to realign the polarization transformation, ensuring it corresponds to measurements in two receivers. The polarization alignment is achieved through QBER minimization using the gradient descent optimization. 

The trend of the measured QBER versus link distance (Fig.~\ref{fig:QKD_Results}) is close to, but slightly higher than the minimum possible value predicted by our PMD model. This observation supports our estimation that, under the given conditions of our QKD system, the PMD-related QBER is dominant and validates the consistency of our model. 

\section{Discussion}

Let us discuss possible ways of PMD mitigation in real-life entangled-based QKD scenarios. First, we look at the simplified model of the first-order PMD approximation. Later in this section, we also evaluate the contribution of higher-order PMD effects to QKD errors.

The overall impact of the PMD-induced loss of fidelity on the QKD system strongly depends on the orientation of the 4-state circle with respect to the PMD vector $\vec{\Omega}$.
For BBM92 protocol (i.e. BB84 bases arrangement) the maximal probability of error occurs when both bases lie on the great circle orthogonal to the PMD vector, Fig.~\ref{fig:spheres}b. Each of the four states experience the same maximally possible decoherence (\ref{eq:model}) at $\varphi = \pi/2$. Detection errors are, thus, equally observed in both bases.

As suggested above, in order to minimize the effect of PMD it is reasonable to align one of the measurement bases with the PMD axis, Fig.~\ref{fig:spheres}c. In this case, the first-order PMD-related decoherence disappears for this particular one, while it remains at the maximal level for the other. Formally, it means that the average PMD-induced QBER is twice as low, as in the worst case.
However, from a practical point of view it creates a strong asymmetry between the two bases, which is not always desirable, as it has to be properly accounted during error correction and privacy amplification steps.

A more practical alternative orientation of the bases is when the PMD vector lies in the plane of the 4-state circle, but exactly in the middle of the arc between two adjacent measurement states, Fig.~\ref{fig:spheres}d.
Due to the factor of $\sin^2\varphi$ in (\ref{eq:model}), the effective probability of errors is still twice smaller than in the worst case, but the errors are evenly distributed between the two bases, which could lead to a more balanced operation and, potentially, better overall performance.

Moreover, if the protocol assumes unequal probabilities for the basis choice, as in ~\cite{Lucamarini2013}, the resulting error distribution can be fine-tuned to achieve the optimal balance of the error rates with respect to the basis choice asymmetry. In other words, the more probable basis should have a lower QBER by shifting the corresponding states closer to the PMD vector while keeping the latter within the 4-state circle plane.

Interestingly, the average detection error probability for the 6-state protocol~\cite{bruss1998sixstate} with equiprobable bases is completely invariant to the rotations of the bases with respect to the PMD vector. This pinpoints the greater symmetry of the 6-state protocol with respect to the conventional BB84. The distribution of errors between the three bases is, however, dependent on the bases orientation and can be easily symmetrized.  

\begin{figure}[ht!]
\centering\includegraphics[width=12cm]{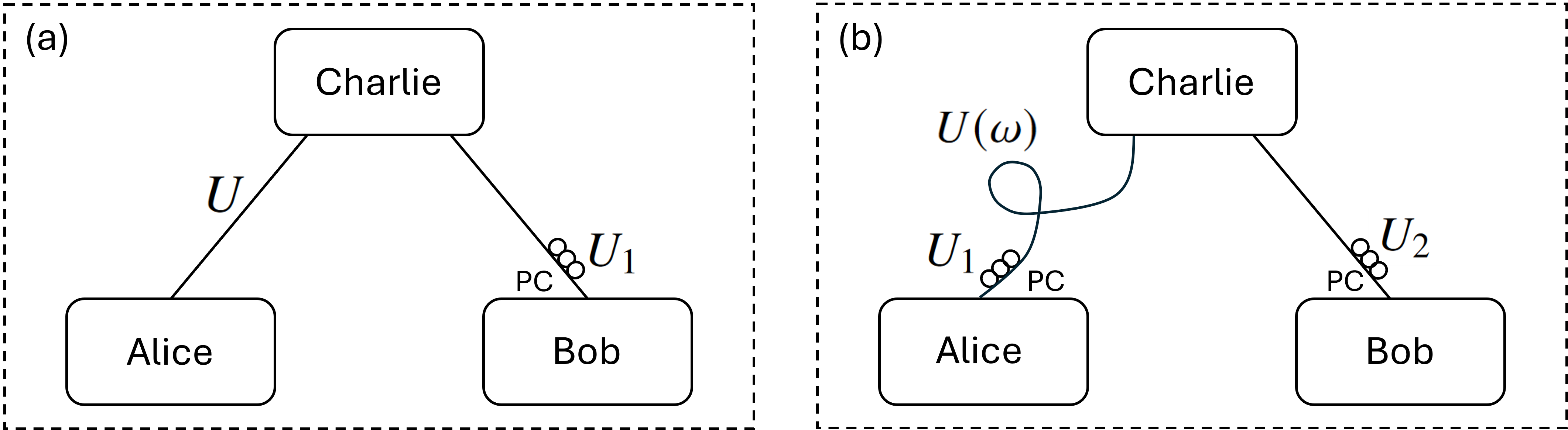}
\caption{Configurations with PMD mitigation: (a) local detection; (b) one local one remote measurement. PC -- polarization controller, $U$ -- unitary transformations of the polarization state.}
\label{fig:PCs_usage}
\end{figure}

Let us now look at a few practical implementation examples to show available tools for first-order PMD mitigation.
For simplicity let us assume that the source generates a singlet state $\bigl(|HV\rangle - |VH\rangle\bigr)/\sqrt{2}$, which can be obtained from the state (\ref{eq:state}) by applying Pauli transformation $\sigma_y$ to either of the qubits. The singlet state, being measured in any basis, always yields perfectly anti-correlated results. Therefore, for establishing QKD it  is only required to match two transformations between the source and the two receivers. This can be realized with a single polarization controller, as shown in Fig.~\ref{fig:PCs_usage}a. It has to be set to perform the same transformation $U$, as in the other channel, so $U_1 = U$.
However, if there is PMD in one of the channels, it is important to properly orient the measurement bases with respect to the PMD vector. As we show below, this can be done by introducing one more polarization controller.

This configuration is shown in Fig.~\ref{fig:PCs_usage}b: one of the photons is detected locally, while the other one passes through the long fiber line with a non-negligible PMD that we denote as a wavelength-dependent unitary polarization transformation $U(\omega)$.
For simplicity, we will consider the case when one of the bases is aligned with the PMD vector.
This can be done with the use of the Alice's polarization controller placed between the channel and her measurement device. It should be tuned to perform the unitary transformation
\begin{equation}\label{eq_unitary1}
U_1 = |H\rangle\langle p_1| + |V\rangle\langle p_2|,
\end{equation} where $\langle p_{1,2}|$ are the two PSPs.

Now a second polarization controller is required to ensure identical transformations in both channels at some median wavelength $\omega_0$. It has to be tuned to perform $U_2 = U_1 U(\omega_0)$. In this configuration the PMD does not affect correlations in the $|H\rangle$--$|V\rangle$ basis ans thus, reduces its overall impact. By the choice of~(\ref{eq_unitary1}) it can be further tuned to become symmetric between the two bases or even to shifting most of errors towards the $|H\rangle$--$|V\rangle$ basis, when it becomes farther from the channel PSPs.

The situation may become even more interesting when two long channels between Charlie and Alice and Bob have their own PMDs. Results provided in the present work give a good insight into available tools of PMD mitigation, which could be easily extended to this more complicated case. Also, there is a known phenomenon of non-local PMD compensation proposed in~\cite{Shtaif2011} and further elaborated in~\cite{Riccardi2020}.
It it is an interesting task to find practical use cases of this compensation concept in  experimental entanglement-based QKD applications. However, it goes beyond the scope of the current manuscript and should be addressed in a separate research initiative.

Now, let us assess the significance of higher-order PMD effects in relation to their contribution to infidelity. First, our experimental results (Fig.~\ref{fig:QKD_Results}) indicate that the overall QKD errors closely align with those predicted by the first-order PMD model. Therefore, there is little scope for higher-order PMD effects to contribute significantly. This leads us to the preliminary conclusion that higher-order PMD effects are less critical to infidelity compared to the first-order effects. 

To assess it numerically, both the absolute value and direction of PMD vector $\vec{\Omega}$ must be examined as they vary with wavelength. Changes in the absolute value indicate variations in the speed at which polarization rotates around $\vec{\Omega}$. Meanwhile, directional changes in $\vec{\Omega}$ lead to deviations of the four measurement states from the plane of the 4-state circle. An explanation of the estimation methods is included in the SM, Section 4. Below, we summarize the key considerations and present the results.

For changes in the absolute value of $\vec{\Omega}$, we notice that the DGD shows oscillations with a period comparable to 1 nm, Fig.~\ref{fig:DGD}. This behavior is typical, as similar oscillations are observed in all our measurements and in \cite{Schiano2004}, for instance. Taking it as a basis for a qualitative model, we conclude that this contribution to the measurement error probability is at least an order of magnitude smaller than the first-order contribution.

We assess the contribution of $\vec{\Omega}$ rotation to the infidelity by evaluating two specific types of polarization states (see SM). For the type 1 state the infidelity is solely due to $\vec{\Omega}$ rotation. In case of the type 2 state the first-order mechanism is maximal. Then, we compare the infidelity for the type 1 state with that for the type 2 state. The result of this comparison shows that the contribution of $\vec{\Omega}$ rotation to the infidelity is an order of magnitude smaller than that given by the first-order mechanism. Interestingly, the contribution from $\vec{\Omega}$ rotation can even be beneficial for PMD mitigation, provided the DGD remains sufficiently small (see SM, very end).

In conclusion, while higher-order PMD effects contribute to infidelity, their impact is significantly smaller than that of the first-order effects.

\section{Conclusion}
For QKD optimization, many parameters are used, such as source brightness, channel losses, dark count rate, detection efficiency, and others. These are engaged in optimization models \cite{ma2005, tayduganov2021} to maximize key generation rates after the last post-processing stage -- privacy amplification. We have expanded these models to take into account PMD-related channel parameters and optimize the filtering bandwidth of broad-band entangled photon sources. 

Our model, validated through experimental measurements, shows a quadratic relationship between infidelity and filtering bandwidth, as well as a linear relationship between infidelity and QKD distance. A simple way to halve the effects of PMD is to align the measurement bases with respect to the PMD vector in such a way that the PMD vector lies in the 4-state circle plane. This can be achieved using an additional polarization controller in the scheme, minimizing QBER. At the same time, it is possible to achieve a balanced error distribution in the measurement bases.

More detailed experiments, as well as further PMD mitigation exploiting non-local PMD compensation, are left for future work. By understanding and mitigating PMD effects, we can improve the security and reliability of QKD systems. We believe our contributions can potentially benefit QKD, quantum communications, and fiber-optic data transfer.

\section{Backmatter}
\begin{backmatter}
\bmsection{Funding}
This project is funded by Abu Dhabi’s Advanced Technology Research Council.

\bmsection{Acknowledgments}
We thank Karen Sloyan, Jaideep Singh, Anton Trushechkin, Rodrigo Sebastian Piera and Vlad Revici for support in various aspects of the QKD system and Juan Villegas for support in early stages of our PMD measurements.

\bmsection{Disclosures}
The authors declare no conflicts of interest.

\bmsection{Data availability} Data underlying the results presented in this paper are not publicly available at this time but may be obtained from the authors upon reasonable request.

\bmsection{Supplemental document}
See Supplement 1 for supporting content. 

\end{backmatter}

%%%%%%%%%%%%%%%%%%%%%%% References %%%%%%%%%%%%%%%%%%%%%%%%%
\bibliography{main}

\end{document}